\documentclass[twocolumn,prl,showpacs,preprintnumbers,amsmath,amssymb,epsf,superscriptaddress]{revtex4-1}

\usepackage{graphicx,epsfig}
\usepackage{dcolumn}
\usepackage{bm}

\begin{document}


\title{Origin and Impacts of the First Cosmic Rays}
\author{Yutaka Ohira}
\affiliation{Department of Earth and Planetary Science, The University of Tokyo, 7-3-1 Hongo, Bunkyo-ku, Tokyo 113-0033, Japan}

\author{Kohta Murase}
\affiliation{Department of Physics, The Pennsylvania State University, University Park, Pennsylvania 16802, USA}
\affiliation{Department of Astronomy \& Astrophysics, The Pennsylvania State University, University Park, Pennsylvania 16802, USA}
\affiliation{Center for Particle and Gravitational Astrophysics, The Pennsylvania State University, University Park, Pennsylvania 16802, USA}
\affiliation{Center for Gravitational Physics, Yukawa Institute for Theoretical Physics, Kyoto, Kyoto 606-8502 Japan}


\begin{abstract}
Nonthermal phenomena are ubiquitous in the Universe, and cosmic rays (CRs) play various roles in different environments. When, where, and how CRs are first generated since the Big Bang?
We argue that blast waves from the first cosmic explosions at $z\sim20$ lead to Weibel mediated nonrelativistic shocks and CRs can be generated by the diffusive shock acceleration mechanism. 
We show that protons are accelerated at least up to sub-GeV energies, and the fast velocity component of supernova ejecta is likely to allow CRs to achieve a few GeV in energy. 
We discuss other possible accelerators of the first CRs, including accretion shocks due to the cosmological structure formation. 
These CRs can play various roles in the early universe, such as the ionization and heating of gas, the generation of magnetic fields, and feedbacks on the galaxy formation.  

\end{abstract}

\maketitle
\section{Introduction}
In the current universe, high-energy nonthermal particles are ubiquitous at various scales from the earth to clusters of galaxies. 
Cosmic rays (CRs) provide one of the best examples. 
The energy density of CRs is about $1\ {\rm eV}~{\rm cm}^{-3}$ in our Galaxy, which is comparable to that of thermal particles. 
Therefore, CRs are thought to have important roles in galaxies. 
It is widely accepted that blast waves of supernova remnants are the origin of CRs with energies up to $10^{15}{\rm eV}$ 
and the standard acceleration mechanism is diffusive shock acceleration (DSA) \cite{axford77}. 
In fact, radio, x-ray and gamma-ray observations of supernova remnants have shown the evidence that electrons and ions are accelerated to highly relativistic energies \cite{koyama95}.  
Furthermore, a CR precursor ahead of the shock front, which is a prediction of the diffusive shock acceleration, was directly imaged \cite{ohira17}.
The CRs could even affect outflow dynamics of galaxies in the halo region, and their energy density is comparable to that of the warm-hot intergalactic medium \cite{mf18}.

It has not been studied when, where, and how the first population of CRs are generated since the Big Bang. 
These questions are, in other words, relevant for us to reveal the history of the nonthermal Universe and their roles in high-redshift environments. 
The CRs in the early universe can heat the intergalactic medium, and future HI 21-cm observations could shed light on the beginning of the nonthermal universe \cite{sazonov15}. 
Ref. \cite{sazonov15} showed that CRs with energies of a few tens MeV most efficiently heat the intergalactic medium, but so far, we do not understand whether the first CRs can be accelerated to the energy scale or not. 
In addition, in order for CRs to heat the intergalactic medium, CRs have to escape from the acceleration site. 
In early universe, there are two energetic shocks that could accelerate the first population of CRs, supernova blast waves of the first stars and accretion shocks of the large scale structure formation in the universe. 
In this work, we consider particle accelerations by the two types of shocks, showing that the first CRs are provided by Weibel mediated nonrelativistic collisionless shocks driven by the first star explosion at $z\approx 20$. 
We use the notation $Q_{a,x}=Q_a/10^x$ in CGS units. 

\section{Supernova blast waves of the first stars and collisionless shocks}
Several cosmological simulations show that the first stars can be formed in dark matter halos with halo masses of $M_{\rm h}\sim 10^6\ M_{\odot}$ at $z\approx 20$ \cite{yoshida03}. 
The first stars are expected to be more massive than the Sun \cite{omukai98} but may have a wide range of masses, $10\ M_{\odot}\lesssim M_* \lesssim 10^3\  M_{\odot}$\cite{hirano14}. 
Because their lifetime is shorter than the Hubble time at the age of the Universe,
the first stars gravitationally collapse at $z\sim5- 20$. 
During their lives, the first stars emit a lot of ultraviolet photons ionizing surrounding gas, forming HII regions. 
The number density, temperature, and ionization fraction of the HII region are typically $n\sim1\ {\rm cm}^{-3}, T\sim1\ {\rm eV}$, and $f_{\rm i}=1$, respectively, just before the first stars end their lives \cite{kitayama04}.

After the core collapse of the first stars, their envelopes may be ejected as supernovae. 
According to a recent cosmological simulation \cite{hirano14}, about $30\%$ of the first stars may directly collapse to black holes and the remaining $\sim70\%$ explode as normal core-collapse, pair-instability, and pulsational  pair-instability supernovae. 
In this work, we mainly consider normal core-collapse supernovae from the first stars as the sources of the first CRs, and the others are discussed later. 

Supernova ejecta consist of the inner core with a shallow density profile and the outer envelope with a steep density profile. 
The bulk of the ejecta has a velocity of $u_{\rm ej}\simeq 4.5 \times10^8\ {\rm cm\ s}^{-1}\ {\mathcal E}_{\rm SN,51}^{1/2}M_{\rm ej,34}^{-1/2}$, where ${\mathcal E}_{\rm SN}$ and $M_{\rm ej}$ are explosion energy and ejected mass, respectively. 
The shock velocity is almost constant, $u_{\rm sh}\approx u_{\rm ej}$, until the swept-up mass becomes comparable to the ejecta mass, $t\lesssim t_{\rm dec}=(3M_{\rm ej}/4\pi m_{\rm p}n)^{1/3}/u_{\rm ej}$, where $m_{\rm p}$ and $n$ are the proton mass and number density in the HII region. 
However, in the earliest phase of the SN expansion, the outer region of the ejecta drives a faster shock with $u_{\rm sh}\gg u_{\rm ej}$ \cite{self}. 
Details depend on mass-loss mechanisms of the first stars, which could be caused by the pulsational instability and perhaps rotationally induced chemical mixing \cite{baraffe01}.  
Assuming a wind density profile of $\varrho_{\rm cs}=DR^{-2}$ with an outer ejecta profile $\varrho_{\rm ej}\propto t^{-3}{(R/t)}^{-\delta} $ (where $\delta=10$ for a radiative envelope with a convective core), we obtain $u_{\rm sh}\simeq 3.3 \times{10}^9\ {\rm cm\ s}^{-1}\ {\mathcal E}_{\rm SN,51}^{7/16}M_{\rm ej,34}^{-5/16}D_{12}^{-1/8}t_{4}^{-1/8}$ \cite{self}. Here we take a fiducial value for blue supergiants, $D\sim10^{12}~{\rm g\ cm}^{-1}$ \cite{smith}. 
If the mass loss is small, the fast shock initially propagates in the wind region, 
and subsequently propagate in the the HII region. 
Thereafter, the transition to the homologous expansion occurs, and finally after $t > t_{\rm dec}$, the shock velocity decreases with time as $u_{\rm sh}\propto t^{-3/5}$. 

To understand whether particles are accelerated by the shock or not, we first investigate what types of collsionless shock and magnetic field turbulence are generated. 
Magnetic fields in the surrounding HII region of the first stars are poorly constrained. 
The Biermann battery effect around the first stars generates magnetic fields of $\sim 10^{-17}\ {\rm G}$ in the $10^2 -10^3\ {\rm kpc}$ scales at $z\approx20$ \cite{doi11}. 
Since the field strength is too small to affect the collisionless shock, the HII region can be treated as an unmagnetized plasma. 
Then, some of downstream hot plasma leak to the shock upstream region. 
Collisionless plasma instabilities are excited between the upstream plasma and the leaking plasma. 
As a result, the upstream cold flow is dissipated by electromagnetic fields generated by the collisionless plasma instabilities and a collisionless shock is formed. 
Note that the shock propagating inside a star is radiation mediated rather than collisionless, and the radiation pressure is important even after the shock breakout. However, collisionless shocks should eventually form when the shock propagates in a wind environment \cite{breakout}, and then the first and earliest CRs will be produced. Then, collisionless shocks of supernova remnants will propagate in the HII region. 

Since numerical simulations show that unmagnetized shocks with $\beta_{\rm sh}=u_{\rm sh}/c\gtrsim 10^{-1}$ are the nonrelativistic Weibel mediated shock \cite{kato08}, the shock driven by the fast-envelope ejecta is so. 
However, unmagnetized shocks with  $\beta_{\rm sh}\approx 10^{-2}$ are poorly understood. 
As a first step, we consider evolution of two counter streaming electron-proton plasmas where the relative velocity is $\beta_{\rm sh} \sim 10^{-2}$ and the temperature of each plasma is $T\sim1\ {\rm eV}$. 
Since the electron mass is smaller than the proton mass, instabilities caused by the two electron beams grow initially. 
If the relative velocity (or shock velocity) is a nonrelativistic velocity,  
growth rates of the electron and ion Weibel instabilities are given by $\gamma \sim \beta_{\rm sh}\omega_{\rm pe}$ and $\gamma \sim \beta_{\rm sh}\omega_{\rm pp}$ \cite{davidson72}, 
and the growth rate of the electron two-stream instability is given by $\gamma \sim \omega_{\rm pe}$ \cite{ichimaru73}. 
$\omega_{\rm pe}$ and $\omega_{\rm pp}$ are the electron and proton plasma frequencies. 
Since the electron two-stream instability is the most unstable mode, the two electron beams initially excite electrostatic fields. 
If only protons leak to the upstream region from the downstream region, 
instead of the electron two-stream instability, 
the Buneman instability becomes the most unstable mode and generates electrostatic fields \cite{buneman58}. 
As a result, only electrons are heated,  
but protons remain cold beams because the frequency of excited waves is close to the electron plasma frequency. 
The initial electrostatic instabilities are saturated when the electron thermal velocity becomes comparable to the relative velocity between the two beams. 
Thus, the electrons are heated to $T_{\rm e}\sim m_{\rm e} u_{\rm sh}^2$, where $m_{\rm e}$ is the electron mass \cite{ohira07}.
In a such plasma, the ion acoustic instability, ion-ion two stream instability, and the ion Weibel instability are unstable. 
Since the ion-ion two stream instability has the largest growth rate of $\gamma \sim \omega_{\rm pp}$ \cite{ohira07}, 
the electrostatic fluctuations are excited and protons are heated to $T_{\rm p} \sim T_{\rm e}\sim m_{\rm e} u_{\rm sh}^2$ \cite{ohira07}. 
Thereby, the ion acoustic instability and the ion-ion two stream instability are stabilized but the ion Weibel instability is still unstable. 
Most of the kinetic energy of the proton beams are not dissipated by the early electrostatic instabilities. 
Then, the ion Weibel instability finally generates magnetic field fluctuations and dissipates the proton beams. 
Therefore, collisionless shocks driven by the core ejecta is also nonrelativistic Weibel mediated shocks.

\section{Acceleration of the first CRs}
{\it Ab-initio} plasma simulations have shown that relativistic Weibel mediated shocks can accelerate particles by DSA \cite{spitkovsky08}. 
Therefore, we can expect that first supernova remnant shocks first accelerate CRs at $z\approx 20$. 
We here derive the acceleration time scale of DSA in the nonrelativistic Weibel mediated shock. 
The acceleration time scale of DSA is given by $t_{\rm acc}=20{\kappa}/u_{\rm sh}^2$ \cite{drury83}, where $\kappa$ is the diffusion coefficient of the accelerated particle, and we assume that the downstream diffusion coefficient is the same as the upstream one. 
The ion Weibel instability generates magnetic field fluctuations, $\delta B$, with the coherent length scale of the proton inertial length, $\lambda_{\rm \delta B}=\alpha c/\omega_{\rm pp}$, where $\alpha$ is a numerical factor.  
The nonlinear evolution of the Weibel instability has been widely discussed theoretically and investigated by numerical simulations \cite{kato05}. Very recently, it is shown that the drift kink instability at $\lambda_{\delta B}\approx 10\ c/\omega_{\rm pp}$ stops the nonlinear growth of the magnetic field fluctuations \cite{ruyer18}. In this work, we adopt $\alpha=10$ as a fiducial value. 
 
The magnetic field strength around the nonrelativistic collisionless shock is subject to more uncertainties. 
To estimate the acceleration timescale, we introduce a magnetic field energy fraction, $\epsilon_{\rm B} = \delta B^2 / (4\pi n m_{\rm p} u_{\rm sh}^2)$. 
The gyroradius of particles with a momentum of $p$ is represented by $r_{\rm g}=  \lambda_{\rm \delta B}\hat{p} /(\alpha \beta_{\rm sh}\epsilon_{\rm B}^{1/2})$, where $\hat{p} = p / (m_{\rm p}c)$. 
If the momentum is larger than $m_{\rm p}u_{\rm sh}\alpha \epsilon_{\rm B}^{1/2}$, the gyroradius is larger than the coherent length scale of magnetic field fluctuations. 
For Weibel mediated shocks, because $\epsilon_{\rm B}$ is smaller than about $10^{-2}$ as argued later, most protons thermalized by the shock satisfy this condition. Then, the diffusion coefficient is given by $\kappa\approx 2\pi vr_{\rm g}^2/\lambda_{\rm \delta B}$ \cite{shalchi09}, where $v$ is the particle velocity. 
Then, the acceleration time scale of DSA in the nonrelativistic Weibel mediated shock is given by
\begin{equation}
t_{{\rm acc},s}=\frac{40\pi}{\alpha \epsilon_{\rm B}\beta_{\rm sh}^4}\frac{\hat{p}^3}{\sqrt{\hat{p}^2+(m_s/m_{\rm p})^2}}\omega_{\rm pp}^{-1}~~,
\label{eq:tacc}
\end{equation}
where the subscript $s={\rm e, p}$ represents particle species. 
The diffusion length around the shock, $\kappa/u_{\rm sh}$, depends on the energy of accelerated particles. 
We naively expect that $\epsilon_{\rm B}$ is not constant in space 
because the number flux density of leaking protons, $f_{\rm leak}$, decreases with the distance from the shock and consequently the electric current that generates magnetic field fluctuations decreases as well as $f_{\rm leak}$. 
From the Biot-Savart law, one can estimate the magnetic field strength by $\delta B / \lambda_{\rm \delta B} \approx 4\pi e f_{\rm leak}/c$. 
In order to estimate the number flux density of protons, $f_{\rm leak}$, we introduce the CR proton energy flux fraction, $\eta_{\rm p} = f_{\rm leak} (\gamma_{\rm p} - 1 )m_{\rm p}c^2 / (nm_{\rm p}u_{\rm sh}^3/2 )$, where $\gamma_{\rm p}$ is the Lorentz factor of leaking protons. 

In the current universe, $\eta_{\rm p}$ is expected to be about 0.1 to supply Galactic CRs by supernova remnants  \cite{baade34}. 
Then, the magnetic field energy fraction in the diffusion region of protons with the momentum of $\hat{p}=(\gamma_{\rm p}^2-1)^{1/2}$ is represented by $\epsilon_{\rm B} = \left\{ \alpha f_{\rm leak} / (nu_{\rm sh}) \right\}^2 = \alpha^2\eta_{\rm p}^2\beta_{\rm sh}^4 / \{4(\gamma_{\rm p} -1)^2\}$. 
The above estimate is based on the CR current in the upstream region. 
The magnetic field fluctuations in the downstream region of all types of collisionless shocks are open issues \cite{keshet}. We simply assume that the diffusion coefficient in the downstream is less than or equal to that in the upstream region. 
This is a reasonable assumption that provides the upper limit of the maximum energy, and has been commonly used in many studies.

For $\hat{p}\gg1$, since electrons and protons have the same diffusion length, and 
$\epsilon_{\rm B}\approx \alpha^2 \eta_{\rm p}^2\beta_{\rm sh}^4\hat{p}^{-2}/4$, 
the acceleration time scale is 
\begin{equation}
t_{{\rm acc},s}=\frac{160\pi \hat{p}^4}{\alpha^3\eta_{\rm p}^2 \beta_{\rm sh}^8} \omega_{\rm pp}^{-1}~~,
\label{eq:taccr}
\end{equation}
which does not depend on particle species. 
For $\hat{p}\ll1$, electrons and protons have the different diffusion length even though they have the same momentum. For nonrelativistic protons, $\epsilon_{\rm B}\approx \alpha^2 \eta_{\rm p}^2\beta_{\rm sh}^4\hat{p}^{-4}$, and the acceleration time scale is
\begin{equation}
t_{\rm acc, p}\approx \frac{40\pi \hat{p}^7}{\alpha^3\eta_{\rm p}^2 \beta_{\rm sh}^8} \omega_{\rm pp}^{-1}~~.
\label{eq:taccn}
\end{equation}
For relativistic electrons with a momentum of $m_{\rm e}/m_{\rm p}\ll \hat{p}\ll 1$, 
since the diffusion length of electrons with a momentum of $\hat{p}$ is that of protons with a momentum of $\hat{p}^{2/3}$,  $\epsilon_{\rm B}\approx \alpha^2 \eta_{\rm p}^2 \beta_{\rm sh}^4\hat{p}^{-8/3}$ and the acceleration time scale is
\begin{equation}
t_{\rm acc, e}\approx \frac{40\pi \hat{p}^{14/3}}{\alpha^3\eta_{\rm p}^2 \beta_{\rm sh}^8} \omega_{\rm pp}^{-1}~~.
\label{eq:tacce}
\end{equation}

The maximum energy of accelerated particles are decided by a finite acceleration time, a finite size of the acceleration region, or cooling \cite{ohira12}. 
Since the acceleration time scale for Weibel mediated shocks strongly depends on the shock velocity, $t_{\rm acc}\propto u_{\rm sh}^{-8}$, 
particles are most efficiently accelerated during the free expansion phase $t\lesssim t_{\rm dec}$. 
The time-limited maximum energy is obtained from the condition $t_{{\rm acc},s}=t_{\rm dec}$, which is almost the same as the size-limited maximum energy for the case of supernova remnants at $t=t_{\rm dec}$ \cite{ohira12}. 
At the deceleration time of the core ejecta, the size-limited maximum energies are given by 
\begin{eqnarray}
E_{\rm max,p}^{\rm dec}&\simeq& 110\ {\rm MeV}\ \alpha_{,1}^{6/7}\eta_{\rm p,-1}^{4/7}M_{\rm ej,34}^{-19/21}{\mathcal E}_{\rm SN,51}n_{,0}^{1/21}~, 
\label{emaxps}
\\
E_{\rm max,e}^{\rm dec} &\simeq& 320\ {\rm MeV}\ \alpha_{,1}^{9/14}\eta_{\rm p,-1}^{3/7}M_{\rm ej,34}^{-19/28}{\mathcal E}_{\rm SN,51}^{3/4}n_{,0}^{1/28}~.
\label{emaxes}
\end{eqnarray}
For the above parameters, only electrons are accelerated to the relativistic energy, and cooling is negligible at $t=t_{\rm dec}$. 

For earlier shocks driven by fast velocity components of the ejecta, the size or time-limited maximum energies of protons are given by 
\begin{equation}
E_{{\rm max},s}^{\rm env} \simeq3.0\ {\rm GeV}\ \alpha_{,1}^{3/4}\eta_{\rm p,-1}^{1/2}M_{\rm ej,34}^{-35/64}{\mathcal E}_{\rm SN,51}^{49/64}D_{\rm 12}^{-3/32}t_{4}^{-7/32}~~.
\label{eq:emaxr}
\end{equation}
For the above parameters, both protons and electrons are accelerated to the relativistic energies, and cooling is negligible. 

Therefore, for first core-collapse supernovae with above parameters, 
CRs are accelerated to a few GeV and a few hundreds of MeV by the fast-envelope and core ejecta, respectively. 
Since the shock velocity decreases with time after the deceleration time of the core ejecta, 
the size or time-limited maximum energy also decreases with time. 
The time evolution is given by 
\begin{eqnarray}
E_{\rm max,p} &=&E_{\rm max,p}^{\rm dec}(t/t_{\rm dec})^{-38/35}~~, \\ 
E_{\rm max,e} &=&E_{\rm max,e}^{\rm dec}(t/t_{\rm dec})^{-57/70}~~. 
\end{eqnarray}

The cooling time of protons due to Coulomb losses is given by $t_{\rm cool}\approx1.5\times10^{15}\ {\rm sec}\ n_{,0}^{-1}\hat{p}^3$. 
For the parameters adopted at Eqs. (\ref{emaxps}) and (\ref{emaxes}), 
the maximum energy of protons is limited by the Coulomb loss for $t\ge t_{\rm c} \simeq 29\ t_{\rm dec}$ 
and the time evolution is given by $E_{\rm max,p}\sim 2.9\ {\rm MeV}\ (t/t_{\rm c})^{-12/5}$. 
Once the maximum energy is limited by cooling ($t>t_{\rm c}$), 
newly accelerated particles cannot escape from the acceleration site. 
Therefore, the first supernova remnants can scatter CR protons in the energy ranges $3\ {\rm MeV}\lesssim E\lesssim3\ {\rm GeV}$ to the intergalactic medium as long as the downstream diffusion coefficient is not much larger than the upstream one. 
They emit gamma rays and produce light nuclei by nuclear interactions, which would affect something and be observable \cite{reeves70}.  

Since magnetic field fluctuations are generated by the Weibel instability driven by the accelerated protons, 
the size of the acceleration region is limited by the diffusion length of protons with the maximum energy, $\kappa/u_{\rm sh}$. 
For $t\ge t_{\rm c}$, the maximum energy of electrons is limited not by the Coulomb loss but by the size of the diffusion region. 
In that case, the maximum energy of electrons is obtained from the condition, $\hat{p}_{\rm max,e}^2=\hat{p}_{\rm max,p}^3$, and one can obtain $E_{\rm max,e}\sim21\ {\rm MeV}\ (t/t_{\rm c})^{-9/5}$. 
All the electrons accelerated by the first supernova remnants can escape from the first supernova remnants and heat the intergalactic medium. 

\section{Other possible accelerators of the first CRs}
Here, we discuss the CR acceleration by explosion phenomena except the normal core-collapse supernova.
For pulsational pair-instability supernovae, stellar mass of $M_{\rm ej}\sim\ 10^{34}\ {\rm g}$, is ejected with ${\mathcal E}_{\rm SN}\sim10^{51}\ {\rm erg}$ before the collapse. 
The remaining stars with $M\sim 10^{35}\ {\rm g}$ finally explode as core-collapse supernovae with ${\mathcal E}_{\rm SN}\sim 10^{51}\ {\rm erg}$ and $M_{\rm ej}\sim 10^{34}-10^{35}\ {\rm g}$ \cite{woosley17}. 
For pair instability supernovae, all the stellar mass, $M\sim\ 2\times 10^{35}\ {\rm g}$, is ejected with ${\mathcal E}_{\rm SN}\sim 10^{53}\ {\rm erg}$. 
About $30\%$ of first stars are expected to collapse to black holes directly, which can eject a small mass 
because neutrino cooling during the protoneutron star phase makes the gravitational mass of the core small. 
The expected ejecta mass and explosion energy are $M_{\rm ej}\sim 10^{30}\ {\rm g}$ and ${\mathcal E}_{\rm SN}\sim 10^{47}\ {\rm erg}$ for massive stars in the current universe \cite{lovegrove13}. 
From Eqs.(\ref{eq:taccr}) and (\ref{eq:taccn}), the pulsational explosion of pulsational pair-instability supernovae, 
the final explosion of pulsational pair-instability supernovae, pair instability supernovae, and the direct collapse of first stars accelerates CR protons to $\sim130\ {\rm MeV}$, $\sim17-130\ {\rm MeV}$, $\sim1.0\ {\rm GeV}$, and $\sim 57\ {\rm MeV}$, respectively. 
It should be noted that the kinetic energy of some supernova ejecta are dissipated not in the HII region but in the stellar wind and ejecta that might be sufficiently magnetized. 
If so, the shock would be a magnetized collisionless shock and CRs would be accelerated to higher energies like CRs in the current universe. 

If first stars explode as a gamma-ray burst or form black-hole binaries, relativistic jets would be produced, resulting in relativistic collisionless shocks and CR acceleration to higher energies \cite{schneider02}. However, whether the relativistic jets are produced by the collapse of first stars and black-hose binaries in the early universe and whether magnetic fields are sufficiently generated in the stellar wind and ejecta are open issues. Therefore, the maximum energy of first CRs could tell us the magnetic field strength around the first star.

\subsection{Acceleration of CRs associated with the earliest cosmological structure formation}
Large-scale structures naturally serve as CR reservoirs, and accretion shocks have been considered as one of the high-energy accelerators in these environments~\cite{accretion}. 
Strong accretion shocks are expected for local massive clusters, while the shocks are expected to be weaker at higher redshifts. 
The physical situation should largely be different in the early universe because typical halos are smaller and the intergalactic medium is less ionized. 

Halo masses that can collapse at $z=20$ within the 3-sigma fluctuations is about $M_{\rm h}\sim 10^6\ M_{\odot}$ \cite{barkana01}. 
The velocity of the accretion shock is about the virial velocity, $u_{\rm sh}\approx v_{\rm vir} = 6.8\times 10^{5}\ {\rm cm\ s}^{-1}\ M_{\rm h,39}^{1/3}\{(1+z)/20\}^{1/2}$. 
In order to accelerate charged particles by the shock, about a half of matters around the accretion shock has to be ionized \cite{ohira}. 
Since most of matters in the universe is not ionized at $z\approx20$, ionization by radiation from the shocked region is important for the particle acceleration. 
However, the shock velocity have to be larger than about $10^{7}\ {\rm cm/s}$ in order to ionize the upstream matter sufficiently \cite{dopita11}. 
Therefore, accretion shocks due to the cosmological structure formation at $z\approx20$ are unlikely to accelerate particles to high energies. 

At $z\lesssim10$, halos with $M_{\rm h}\sim 10^{10}\ M_{\odot}$ can collapse and the velocity of the accretion shock is about $10^{7}\ {\rm cm\ s}^{-1}$ \cite{barkana01}. 
The ionization fraction and temperature of the shock upstream region are $f_{\rm i}\sim 0.2\ u_{\rm sh,7}^2$ and $T\sim1\ {\rm eV}$ \cite{dopita11}. 
The photoionization precursor length is $10^{20}\ {\rm cm}\ n_{,-1}^{-1}$ \cite{sutherland17}. 
In the spherical top-hat halo model, the baryon number density is given by $n\sim 18\pi^2\ {\bar n_{\rm b}}=4.2\times10^{-2}\ {\rm cm}^{-3}\ \{(1+z)/10\}^3$ when the halo is virialized, where ${\bar n_{\rm b}}$ is the mean baryon number density of universe at the virialization time. 
Magnetic fields in the upstream region of the accretion shock would be amplified by the turbulent dynamo or CR streaming because accretion flows would be very complicated and the first CRs have already accelerated at $z\approx20$. 
However, since the above magnetic-field amplification and generation are still open issues, here we assume that 
the accreting plasma is unmagnetized. 
Then, since the thermal velocity of electrons, $v_{\rm th,e}=4.2\times10^7\ {\rm cm\ s}^{-1}\ (T/1{\rm eV})^{1/2}$, is larger than the shock velocity of the accretion shock, only the Weibel instability is unstable in the collisionless shock transition region. 
Therefore, the accretion shock at $z\lesssim10$ is also the Weibel mediated nonrelativistic shock and 
can accelerate particles by DSA. 
The acceleration time scale is given by Eq.(3) or Eq.(4). 
For the accretion shock at $z\lesssim10$, the maximum energy of accelerated protons is nonrelativistic and 
limited by cooling due to the Coulomb loss. 
Then, the cooling-limited maximum energy is obtained from the condition $t_{\rm acc,p} = t_{\rm cool}$:
\begin{equation}
E_{\rm max,p} \simeq  330\ {\rm keV}\ \alpha_{,1}^{3/2}\eta_{\rm p,-1}\beta_{\rm sh,-3}^4 n_{,-1}^{-1/4}~~.
\end{equation}
Since the acceleration of protons is not limited by the size of acceleration region, most accelerated protons cannot escape to the upstream region. 
Even though the accelerated protons escape to the upstream region of the accretion shock, 
they are quickly thermalized and cannot heat the far upstream region of the accretion shock. 
The free streaming distance during the cooling time of ionization is $\sim2\ {\rm kpc}\ {[(1+z)/21]}^{-3}{(E/330~\rm keV)}^2$ \cite{sazonov15}, 
which is much smaller than that of protons accelerated by the first supernova remnants. 
For electrons, the maximum energy is limited by the size of the diffusion length of accelerated protons, where the cooling due to the ionization loss is negligible for electrons. 
Then, the maximum energy of electrons is obtained from the condition, $\hat{p}_{\rm max,e}^2=\hat{p}_{\rm max,p}^3$: 
\begin{equation}
E_{\rm max,e} \simeq 4.1 \ {\rm MeV}\ \alpha_{,1}^{9/8}\eta_{\rm p,-1}^{3/4}\beta_{\rm sh,-3}^{3} n_{,-1}^{-3/16}~~. \nonumber
\end{equation}
All the accelerated electrons can escape from the accretion shock and heat the intergalactic medium. 

\section{Impacts on cosmological evolution}
Our study has several important implications for the cosmological evolution of galaxies and intergalactic medium. 
First, low-energy CRs may impact the heating and ionization of ambient gas in the early universe. In the current universe, the local CR luminosity density is $Q_{\rm CR}\approx2\times{10}^{46}\ {\rm erg}\ {\rm Mpc}^{-3}{\rm yr}^{-1}({\rm SFR}/0.015\ M_\odot\ {\rm Mpc}^{-3}{\rm yr}^{-1})$\cite{mf18}, where $\rm SFR$ is the star-formation rate density. With a massive POP-III SFR of $\sim{10}^{-5}-{10}^{-4}\ M_\odot\ {\rm Mpc}^{-3}{\rm yr}^{-1}$ around $z\sim20$, we have $Q_{\rm CR}^{\rm III}\sim{10}^{43}-{10}^{44}\ {\rm erg}\ {\rm Mpc}^{-3}{\rm yr}^{-1}$. By comparing the CR loss time $t_{\rm loss}(z)$ to the Hubble time at $z$, $t_{H}(z)\approx1.9\times{10}^{8}{[(1+z)/21]}^{-3/2}\ {\rm yr}$, the CR energy density is estimated to be $u_{\rm CR} \sim3\times10^{-18}\ {\rm erg\ cm^{-3}}{[(1+z)/21]}^{3}({\rm min}[t_H(z),t_{\rm loss}(z)]/100\ {\rm Myr})$ 
$({\rm SFR}_{\rm III}/{10}^{-4}\ M_\odot~{\rm Mpc}^{-3}{\rm yr}^{-1})$ in the physical volume, and the similar value can be obtained by using a POP-III supernova rate of ${10}^{-7}-{10}^{-6}\ {\rm Mpc}^{-3}\ {\rm yr}^{-1}$ and a typical kinetic energy of ${\mathcal E}_{\rm SN}\sim{10}^{51}\ {\rm erg}$.   

A low-energy part of first CRs ($E\lesssim 30\ {\rm MeV}$) ionizes and heats the neutral intergalactic medium. The temperature shift is estimated by~\cite{sazonov15}
\begin{eqnarray}
{\Delta T}_{\rm IGM}&\approx& \frac{2f_{\rm heat}Q_{\rm CR}^{\rm III}t_{H}(z)}{3\Omega_b\varrho_{\rm crit}/m_p}\sim4\ {\rm K}\ \left(\frac{f_{\rm heat}}{0.25}\right){\left[\frac{(1+z)}{21}\right]}^{-3/2}\nonumber\\
&&~~~~~~~~~~~~~~~~~~~\times\left(\frac{{\rm SFR}_{\rm III}}{{10}^{-4}\ M_\odot {\rm Mpc}^{-3}{\rm yr}^{-1}}\right),  
\end{eqnarray}
where $f_{\rm heat}\sim0.25$ is the energy fraction used for heating, $\Omega_b=0.04$ and $\varrho_{\rm crit}\approx1.88h^2\times{10}^{-29}~{\rm g}~{\rm cm}^{-3}$ is the critical density. 
Given that the CMB temperature is $T_{\rm CMB}=2.7~{\rm K}~{(1+z)}$, the effect of the first CRs may be observed by future HI 21-cm observations, especially if the first CR production rate comoving density may continue down to lower redshifts, e.g., $z\approx 6$ (see \cite{desouza11} for the redshift evolution of the POP III SFR). 

Second, the first CRs may provide seed fields for intergalactic magnetic fields. 
In our galaxy, the magnetic field energy density is comparable to the CR energy density. 
If the energy of the first CRs is converted to the energy of magnetic fields by some plasma effects at $z\approx 20$, the comoving magnetic field strength is estimated to be 
\begin{eqnarray}
B_0&=&(1+z)^{-2}\sqrt{\varepsilon_B8\pi u_{\rm CR}}\nonumber\\
      &\sim&3\times{10}^{-11}\ {\rm G}\ \varepsilon_B^{1/2}~{\left[\frac{(1+z)}{21}\right]}^{-5/4}\nonumber\\
&&\times {\left(\frac{{\rm SFR}_{\rm III}}{{10}^{-4}\ M_\odot{\rm Mpc}^{-3}{\rm yr}^{-1}}\right)}^{1/2},  
\end{eqnarray}
where $\varepsilon_B$ is the energy fraction transferred to magnetic fields and the magnetic flux conservation is assumed to estimate $B_0\equiv B(z=0)$. 
Note that the cooling time of CRs above 100 MeV is typically longer than the Hubble time for the average intergalactic medium density. 
This value is consistent with lower bounds on the current intergalactic magnetic fields, $10^{-18} - 10^{-15}\ {\rm G}$ \cite{IGMFreview}, which can be explained if a very small fraction of the first CR energy can be converted the magnetic field energy. Testing such a magnetic field strength is challenging but could be probed by ultrahigh-energy gamma-ray observations \cite{murase09}. 
Moreover, the magnetic field generated by the first CRs at $z\approx 20$ is sufficiently large as a seed for galactic dynamos to explain current galactic magnetic fields~\cite{davis99}, which is especially the case if the first CRs are trapped in protogalaxies. 

Finally, we point out that the first CRs may eventually lead to CR-driven wind once magnetic fields are amplified. It has been suggested that CRs play an important role in the evolution of galaxies \cite{pakmor16}, and such feedback on protogalaxies should be examined in more detail and may be important for cosmological simulations of galaxy formation.

\section{Conclusions}
We studied the origin of the first CRs. We have found the followings: 
1) The first CRs can be accelerated by the first supernova remnants driven by the explosion of first stars at $z\approx20$ (corresponding to about 180 million years after the big bang). 
2) The first CRs are accelerated by DSA in the nonrelativistic Weibel mediated shocks, although CR streaming instabilities can play some roles in the presence of violent mass losses in the preexplosion phase. 
3) The maximum energy of the first CRs lies in the $0.1-10\ {\rm GeV}$ range. As a result, secondary neutrinos produced by these CRs are not expected in the IceCube range unless supernovae are accompanied by choked jets or dense magnetized environments \cite{neutrino}. 
4) The first CR protons with energies larger than a few MeV can escape from the first supernova remnants, whereas all accelerated electrons can. 
These CRs may impact the heating, ionization, magnetic field generation, and the cosmological evolution of protogalaxies. In particular, we suggest that HI 21cm signatures can be used as a powerful test for the first CRs. 
5) Accretion shocks associated with the cosmological structure formation at $z\approx 20$ are unlikely to accelerate particles because the shocks propagate in neutral media. On the other hand, the accretion shocks at $z\lesssim 10$ can accelerate particles. The maximum energies of protons and electrons are a few hundreds keV and several MeV, respectively. Only the accelerated electrons can escape from the accretion shock without energy losses. 

\begin{acknowledgments}
We thank Tom Abel, Katusaki Asano, Takashi Hosokawa, Fumio Takahara, and Ryo Yamazaki for useful comments. 
This work is supported by JSPS KAKENHI Grant Number JP16K17702 (Y.O.), JP19H01893(Y.O.) and Alfred P. Sloan Foundation and NSF Grant No. PHY-1620777 (K.M.).  Y.O. is supported by MEXT/JSPS Leading Initiative for Excellent Young Researchers. 
\end{acknowledgments}

\end{document}